\documentstyle[twoside,fleqn,espcrc2,epsf]{article}

\newcommand{\AmS}{{\protect\the\textfont2
  A\kern-.1667em\lower.5ex\hbox{M}\kern-.125emS}}
\newcommand{\etal}{{\it et al.}}

\title{ Improved flavor symmetry in Kogut-Susskind fermion actions }

\author{K. Orginos,\address{Department of Physics, University of Arizona, Tucson, AZ 85721, USA}        
        \thanks{Presented by K. Orginos. Supported by the US DOE and NSF.
                Computations were done at ORNL, PSC, NCSA, BU, and NERSC 
                as part of the MILC collaboration.}
 R. Sugar, \address{Department of Physics, University of California, Santa Barbara, CA 93106, USA} 
and D. Toussaint$\,\null^{\rm a}$
}
       
\begin{document}

\begin{abstract}
 We study improved Kogut-Susskind fermion actions focusing on flavor
symmetry restoration. Several variants of fat actions suitable for
dynamical simulations are considered, including an action with no tree
level $O(a^2)$ errors.  The spectrum of all the pions is computed and
used as a measure of flavor symmetry violation. Finally, the Naik term
is introduced to restore rotational symmetry.  %RS
\end{abstract}

\maketitle

\section{INTRODUCTION}

Dynamical fermion simulations are still one of the challenging
problems in lattice QCD. Recently, a lot of work has been devoted to %RS
developing improved lattice fermion actions. In this paper, we report
our findings on improved Kogut-Susskind fermions. In particular, we
focus on the issue of flavor symmetry, and the construction of an
action suitable for dynamical simulations.

Kogut-Susskind (KS) fermions have $O(a^2)$ lattice artifacts, in
contrast to the Wilson fermions which have $O(a)$. They represent four
degenerate quark flavors with a $U(4)\times U(4)$ symmetry in the
continuum. This symmetry is broken down to a $U(1)\times U(1)$ on the
lattice. This remnant symmetry protects against additive mass
renormalizations.  The $U(4)\times U(4)$ symmetry breaking manifests
itself in mass splittings in the pion spectrum. There is only one
Goldstone pion associated with the exact $U(1)$ chiral symmetry, while
the rest of the fifteen pions~\cite{GOLTERMAN_MESONS} are
non-Goldstone massive particles.

The first step towards improving the KS action was made by Naik. He
introduced the so called ``Naik term'' ($ -\frac{1}{6}\Delta^3_\mu $),
which eliminates a tree level $O(a^2)$ artifact, improving the
Lorenz symmetry (dispersion relation). The remaining order $O(a^2)$
artifacts are the $U(4)\times U(4)$ breaking terms. As
Lepage~\cite{LEPAGE97} pointed out, these terms are due to couplings
to high momentum gluons, which cause scattering of lattice quarks
among the corners of the Brillouin zone.  These scatterings correspond
to spin-flavor transformations of the physical quark field, causing
the breaking of the symmetry. Thus, the way to reduce the symmetry
breaking is to smear~\cite{MILC_FATLINKS,SL,TD_AH_TK,LEPAGE98} 
the quark-gluon vertex, which suppresses the couplings to high
momentum gluons.  Recently, it was shown by Lepage~\cite{LEPAGE98}
that the one gluon exchange interaction at high momentum is identical
to 4-fermi $U(4)\times U(4)$ breaking terms. Consequently, these
effects can be canceled by either appropriate smearing of the
quark-gluon vertex, so that high momentum gluons decouple, or by 
including in the action appropriate 4-fermi terms.

\section{ACTION CONSTRUCTION}

 Since the $U(4)\times U(4)$ symmetry breaking occurs due to coupling
to high momentum gluons, we need to modify the quark-gluon vertex in
order to decouple them. The coupling of quarks to             %RS
a gluon $A_\mu$ with any of its transverse momentum components
$p_\nu=\pi/a$ causes quark scattering among the corners of the
Brillouin zone.  Couplings to gluons with longitudinal momentum
$p_\mu=\pi/a$ vanish due to cancellations between the forward and the
backward part of the Dirac operator, consequently we do not need to
worry about them.  There are 3 independent couplings to high
transverse momentum gluons ($V_1,V_2,V_3$). We can eliminate them by
modifying the gauge link used in the lattice derivative as following:

\begin{eqnarray}
U_\mu(x)\!\!\!\!&\rightarrow&\!\!\!\! c_1U_\mu(x)+\sum_\nu \Big[ w_3S^{(3)}_{\mu\nu}(x)+\nonumber\\
\!\!\!\!&+&\!\!\!\!\sum_\rho \Big( w_5 S^{(5)}_{\mu\nu\rho}(x) + 
\sum_\sigma w_7 S^{(7)}_{\mu\nu\rho\sigma}(x)\Big)\Big] 
\label{smearing}
\end{eqnarray}
\begin{eqnarray}
\lefteqn{S^{(3)}_{\mu\nu}(x) = U_\nu(x)
         U_\mu(x+\hat\nu)U^\dagger_\nu(x+\hat\mu)}\nonumber\\
\lefteqn{S^{(5)}_{\mu\nu\rho}(x) = U_\nu(x)
         S^{(3)}_{\mu\rho}(x+\hat\nu)U^\dagger_\nu(x+\hat\mu)}\nonumber\\
\lefteqn{S^{(7)}_{\mu\nu\rho\sigma}(x) = U_\nu(x)
         S^{(5)}_{\mu\rho\sigma}(x+\hat\nu)U^\dagger_\nu(x+\hat\mu)}
\label{staples}
\end{eqnarray}
It is not difficult to see that the above set of terms is the simplest
needed to eliminate completely couplings to high momentum gluons.

In the weak coupling limit, the couplings ($V_1$,$V_2$, $V_3$) to the
gauge field with one, two or three
of the transverse momentum components $\pm\pi/a$ can be
expressed as functions of the staple couplings $w_3,w_5,w_7$, and
the single link coupling $c_1$:
\begin{eqnarray}
V_1 &=& c_1 +  2 w_3 -  8 w_5 - 48 w_7  \nonumber\\
V_2 &=& c_1 -  2 w_3 -  8 w_5 + 48 w_7  \nonumber\\
V_3 &=& c_1 -  6 w_3 + 24 w_5 - 48 w_7  
\label{vertex}
\end{eqnarray}
The overall normalization condition
\begin{equation}
c_1  + 6 w_3 + 24 w_5 + 48 w_7 = 1
\label{normalization} 
\end{equation}
is used to ensure that the total coupling to the nearest neighbor in
the free field limit is one.  The solution to the equations
$V_1=0,V_2=0,V_3=0$ is $c_1 = 2 w_3 = 8 w_5 = 48 w_7 = 1/8 $.  This
set of parameters defines our ``Fat7'' action.  The ``Fat5'' action is
constructed by using paths up to length 5, minimizing the maximum of
the couplings $|V|$. The ``Fat5'' couplings are $c_1 = 2 w_3 = 8 w_5 =
1/7$, $w_7=0$, which give $|V_1|=|V_2|=|V_3| = 1/7 $.  Adding the Naik
term to the ``Fat7'' action results in an action with no tree level
$O(a^2)$ errors and improved flavor symmetry. Unfortunately, the extra
couplings introduced to fix the high momentum behavior introduce a low
momentum artifact $\sim A_\mu k_\nu^2 a^2$.  In order to correct this
new artifact, one has to introduce a new term discovered by
Lepage~\cite{LEPAGE98}. This new term is a double staple
\begin{equation}
\lefteqn{S^{(L)}_{\mu\nu}(x) = U_\nu(x)S^{(3)}_{\mu\nu}(x+\hat\nu)
         U^\dagger_\nu(x+\hat\mu)},
\label{lepa}
\end{equation}
with weight $w_L$. Cancellation of the new artifact occurs, if $w_3
\rightarrow w_3 - 6 w_L$ and $w_L=-1/16$. Note that because the new
term couples to a distance 2 in the $\nu$ direction, at momentum
$p_\nu=\pi/a$ it acts as a single link term. For this reason, the
above change of couplings preserves the normalization condition and
keeps the couplings $V_1$, $V_2$, and $V_3$ zero. Adding the Naik
term, ($ -\frac{1}{6}\Delta^3_\mu $) we get an action which we call
``Asq'', that has errors of $O(a^4,g^2a^2)$. All the above actions can
be tadpole improved. For the tadpole improved actions, we append the
suffix ``tad'' to the names above.

Motivated by the success of RG inspired APE smeared actions in the
work of DeGrand, Hasenfratz and collaborators~\cite{TD_AH_TK}, we also
decided to approximate APE smeared actions as a set of paths with
appropriate weights. This is necessary, if we want to use such actions
in dynamical simulations.  In~\cite{OST} we show how this
approximation can be done expanding in powers of the APE smearing
parameter $\alpha$.  For one APE smearing with $\alpha=.25$, and to
leading order in $\alpha$ we get the action we call ``OFUN25'' (once
fattened unitarized $\alpha=.25$). The fat link for this action is:
\begin{equation}
U_\mu\rightarrow U_\mu +\frac{\alpha}{2}\sum_\nu\left(S^{(3)}_{\mu\nu}-U_\mu S^{(3)}\null^\dagger_{\mu\nu}U_\mu\right).
\label{AP_UN}
\end{equation}

\section{SIMULATIONS AND RESULTS}

\begin{figure*}[t]
\epsfxsize=\textwidth
\vspace{-.4cm}
\epsfbox{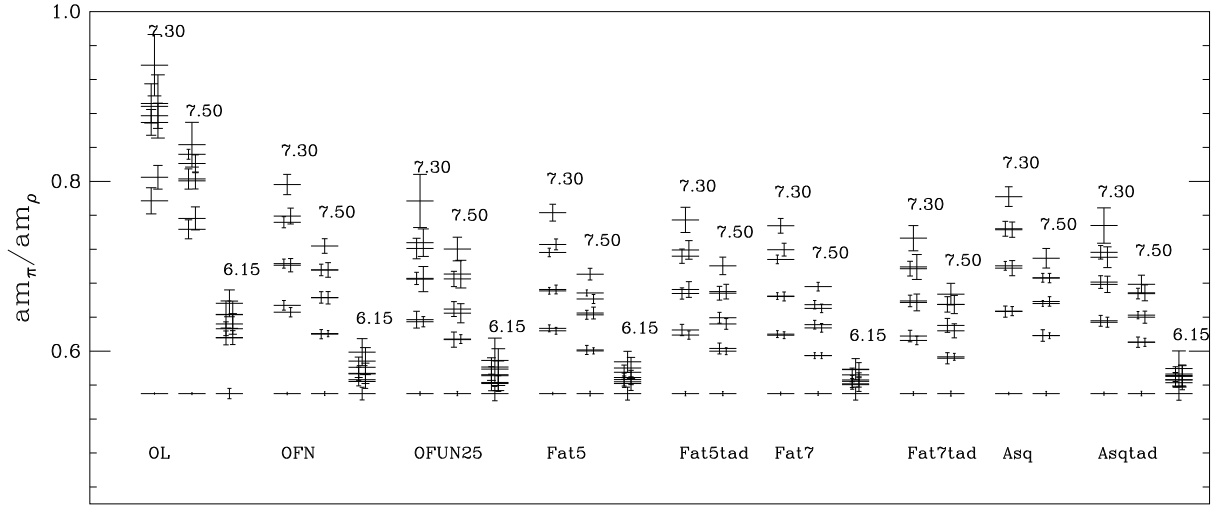}
%\vspace{-.23cm}
\vspace{-.5cm}
\caption{Interpolated pion masses for all the actions tested. The
highest level is the flavor singlet pion while the lower is the
Goldstone pion. The first doublet is the local non-Goldstone pion
(flavor structure $\gamma_0\gamma_5$) and the $\gamma_i\gamma_5$ pion
(right). The second is the $\gamma_0\gamma_i$(left) and
$\gamma_i\gamma_j$ pion. The third is the $\gamma_0$(left) and
$\gamma_i$ pion.}
\vspace{-.18cm}
\label{Figure}
\end{figure*}

 Our action tests are focused on the flavor symmetry restoration,
since we have established in our previous studies~\cite{OT} that the
Naik term improves Lorenz symmetry adequately for the lattice spacings
used in current simulations. Furthermore, the Naik term has little
effect on flavor symmetry.  Thus, in most of the smearing variants we
use, we do not include the Naik term.  We include the Naik term into
``Asq'' and ``Asqtad'', since these actions are the candidates for an
action with both improved flavor and rotational symmetry. The flavor
symmetry breaking is estimated by the splittings in the pion spectrum.
Spectroscopy is done on the same set of lattices (sample sizes range
from 48 to 60 ) for all the actions tested. We use $12^3\times 32$
$\beta= 7.3$ lattices with two dynamical quark masses $m=.020$ and
$m=.040$, and $16^3\times 48$ $\beta=7.5$ lattices with dynamical
quark masses $m=.015$ and $m=.030$.  The gauge action is 1-loop
Symanzik improved and the sea quark action is Kogut-Susskind with a
staple and a Naik term~\cite{OT,OST}. We also use a set of $32^3\times
64$ quenched lattices at $\beta=6.15$ with Wilson gauge action.  The
pion spectra are computed for the two different masses in the case of
the dynamical lattices and for $m=.010$ and $m=0.020$ in the case of
the quenched lattices. We then interpolate the results to  %RS
$m_\pi/m_\rho=.55$.  The interpolated spectra are shown in
Figure~\ref{Figure}. In this figure, we also show for comparison the
spectrum of the standard KS action (OL) and the spectrum of the action
used for the dynamical quarks(OFN).
Our results in Figure~\ref{Figure} show that there is a significant
improvement of the flavor symmetry as we increase the suppression of
the high momentum gluon couplings. The highest improvement is observed
for the ``Fat7tad'' and ``Asqtad'' actions where couplings are set to
zero at tree level. For all the actions the tadpole variant is better
than the non-tadpole. The approximately reunitarized action ``OFUN25''
is slightly better than the ``OFN''.  In all cases the spectrum
exhibits the degeneracies suggested in~\cite{Sharpe}.  For a detailed
error analysis that better justifies our claims see our recent
publication~\cite{OST}.

Although the differences in flavor symmetry improvement among the
various highly smeared actions seem small, we favor ``Asqtad'' on
theoretical grounds ($O(a^4,g^2a^2)$ errors), and on our estimation
that it only costs a factor of 2-3 for dynamical simulations with
light quarks. We believe that such an action will be useful in
realistic projects with dynamical quarks.


\begin{thebibliography}{9}

\bibitem{GOLTERMAN_MESONS}
M.F.L. Golterman, Nucl. Phys. B {\bf 273} (1986) 663.

\bibitem{LEPAGE97}
G.P. Lepage, Nucl. Phys. (Proc. Suppl.) {\bf 60A} (1998) 267. 

\bibitem{MILC_FATLINKS}
T. Blum \etal, Phys. Rev. D {\bf 55} (1997) 1133.

\bibitem{SL} J.F.~Lagae, D.K.~Sinclair, hep-lat/9806014;
             Nucl. Phys. (Proc. Suppl.) {\bf 63} (1998) 892.

\bibitem{TD_AH_TK} %hep-lat/9710078
T.~DeGrand, A.~Hasenfratz and T.~Kovacs,
Phys. Lett. {\bf 420B} (1998) 97.

\bibitem{OT} 
K. Orginos, D. Toussaint 
Phys. Rev. D {\bf 59} (1999) 014501; 
Nucl. Phys. (Proc. Suppl.) {\bf 73} (1999) 909. 

\bibitem{OST} 
K. Orginos, R. Sugar, D. Toussaint Phys. Rev. D {\bf 60 } (1999) 054503.

\bibitem{LEPAGE98} %eprint hep-lat/9809157.
G.P. Lepage, Phys. Rev. D {\bf 59} (1999) 074502.

\bibitem{Sharpe} W. Lee, S. Sharpe hep-lat/9905023.
 
\end{thebibliography}
\end{document}